\documentclass[twocolumn,aps,prl]{revtex4}
\usepackage[english]{babel}
\usepackage[dvipdf]{graphicx}
\usepackage[dvips]{epsfig}
\usepackage{amssymb}
\usepackage{amsmath}
\usepackage{amsbsy}
\usepackage{dcolumn}

\usepackage[section]{placeins}
\usepackage{color}

\begin{document}

\title{Impact of beads and drops on a repellent solid surface: a unified description}

\author{S. Arora}
 \altaffiliation{L2C, University o Montpellier \& CNRS.}
\author{J-M Fromental }
 \altaffiliation{L2C, University o Montpellier \& CNRS.}
\author{S. Mora}
 \altaffiliation{LMGC, University o Montpellier \& CNRS.}
\author{Ty Phou}
 \altaffiliation{L2C, University o Montpellier \& CNRS.}
\author{L. Ramos}
 \altaffiliation{L2C, University o Montpellier \& CNRS.}
\author{C. Ligoure}
 \altaffiliation{L2C, University o Montpellier \& CNRS.}
\email{christian.ligoure@umontpellier.fr}

\affiliation{
 Laboratoire Charles Coulomb (L2C), University of Montpellier, CNRS, Montpellier, France\\
 LMGC,University of  Montpellier, CNRS, Montpellier, France
}

\date{\today}

\begin{abstract}

We investigate freely expanding sheets formed by ultrasoft gel beads, and liquid and viscoelastic drops, produced by the impact of the bead or drop on a silicon wafer covered with a thin layer of liquid nitrogen that suppresses viscous  dissipation thanks to an inverse Leidenfrost effect. Our experiments show a unified behaviour for the impact dynamics that holds for
 solids, liquids, and viscoelastic fluids and that we rationalize by properly taking into account elastocapillary effects. In this framework, the classical impact dynamics of solids and liquids as far as viscous dissipation is negligible, appears as the asymptotic limits of a universal theoretical description. A novel material-dependent characteristic velocity that includes both capillary
and bulk elasticity emerges from this unified description of the physics of impact.

\end{abstract}

\maketitle

Impact of bodies is at the core of a wide range of fundamental and practical areas including, aerosols, erosion, coating, biomechanics, sport, biotechnology... The way in which a liquid drop or an elastic bead deforms during its impact on a solid surface is a daily life fascinating rapid process. It has however eluded explanation until the past $20$ years, when high-speed video technology began to allow time-resolved observations~\cite{Josserand:2016jf}.

Owing to the numerous environmental and industrial applications, the impact of liquid drops on solid surfaces has been studied  extensively since the pioneering work of Worthington~\cite{Worthington1876} till now~\cite{YARIN:2006ft,Josserand:2016jf} and displays extremely diverse and surprising phenomena. The impact may result in the drop spreading over the solid surface, receding, splashing, rebounding, depending on the impact velocity, the  drop size, the properties of the liquid (its density, viscosity, viscoelasticity), the surface and  interfacial tensions, the roughness and wettability of the solid surface. On repellent surfaces (those include superhydrophobic surfaces \cite{Richard2002}, hot plates above the Leidenfrost temperature \cite{Biance2003} or sublimating surfaces \cite{Antonini2013}),  and for high impact velocity $v_0$, a rebound phenomenon is systematically observed~\cite{Antonini2013}. In this case, once the viscous forces are negligible, the spreading dynamics results  solely  from a balance between inertia and capillary forces, which is characterized by the Weber number $We=(\rho  d_0 v_0^2)/\gamma$, where $\rho$, and $\gamma$ are the liquid density and surface tension  respectively. The balance  leads to  a rebound time $\tau_R\propto(\rho d_0^3/\gamma)^{1/2}$  independent of the impact velocity~\cite{Richard2002,Wachters:1966vy} and to a maximum spreading factor, $\lambda_{max}$, defined as the ratio between the diameter of the sheet at its maximal expansion, $d_{max}$, and the diameter of the drop, $d_0$, which is only a function of $We$. The experimentally measured  scaling  $\lambda_{max} \propto We^{0.4}$~\cite{Tran2012, Antonini2013} is close  to the predicted scaling, $\lambda_{max} \propto We^{1/2}$~\cite{Josserand:2016jf}, which is however difficult to observe because of the presence of splash at high $We$.

On the other hand, the impact of an elastic bead on a solid surface has by contrast attracted less attention. Tanaka \textit{et al.}~\cite{Tanaka2003,Tanaka:2005kb} have reported on the impact of compliant solid spherical balls made of  cross-linked gel of centimeter size in non sticking conditions. They have shown that the spreading dynamics can be rationalized from a balance between inertia and bulk  elastic forces and depends on the adimensional Mach number $M=v_0/U_s$, where $U_s= \sqrt{G_0/\rho}$ is the  velocity of transverse  sound waves in the elastic medium and $G_0$ is the shear modulus of the gel. At high impact velocity ($\lambda_{max}\gg1$), a maximum spreading factor $\lambda_{max}\propto M$
and a rebound time independent of the impact velocity $\tau_R\propto(\rho d_0^2/G_0)^{1/2}$ are predicted.

Notably, impact of yield stress fluids reveals either a solid-like behavior~\cite{luu2009drop} or a liquid-like behavior depending on the experimental conditions~\cite{Chen:2016fo}. Despite the intermediate between liquid and solid behavior of some complex fluids, the impact dynamics of liquids~\cite{Richard2002} and solids~\cite{Tanaka2003} have apparently nothing in common even though they seem separately rather well understood.  In this Letter, we show that their behavior can be unified. Here we revisit the impact dynamics of both ultrasoft  elastic beads, and drops of viscoelastic or simple fluids, all with a same millimetric size $d_0$  in  the same experimental conditions such that viscous dissipation and/or solid friction effects  can be safely neglected. The elastic and viscoelastic samples have been carefully chosen, so that the elastocapillary length  $l_{ec}  \equiv 3\gamma/G_0$ lies in the range  $0.1 \times d_0\lesssim l_{ec}\lesssim10 \times d_0$. When $l_{ec} \ll d_0 $, one expects that the deformation of the samples is dominated by  surface tension, whereas in the opposite case  ($l_{ec} \gg d_0$), surface tension effects can be neglected, and the deformations of the samples should  be dominated by bulk elasticity.  In intermediate cases both effects play a role.  Note that for simple liquids  $l_{ec}=0$. The importance of considering the surface  energy for  the mechanics of soft materials is indeed an emerging field \cite{Style:2017ji} that has been recently highlighted  for instance in the framework  of composite materials\cite{Style:2014gv} , wetting \cite{Nadermann:bs,Schulman:2015kp} or adhesion\cite{Chakrabarti:2013dc} phenomena.

  Here we reveal, for the first time to best of our knowledge,  a coupling of  elasticity and surface tension of solids with dynamics. Moreover, the finite deformations that occur in our experiments are definitively far larger than those involved in the previous studies of elasto-capillarity. We describe the dynamics of sheets that result from the impact of  various classes of samples on a repellent surface. We focus more particularly on the maximum spreading factor and  on the time to reach the maximum size of the sheet (roughly half of the rebound  time) as a function of the impact velocity. We show that the previous models  used to describe  the impact of solids (respectively liquids) do not hold for ultrasoft beads (respectively viscoelastic drops). By including elastocapillary effects in these models, we obtain an excellent agreement between theory and experiments for solids and viscoelastic liquids. A new material-dependent  velocity characteristic of the  generalized elastic deformations of materials emerges from this unified description, which we show to be also valid for the impact of simple liquids.

To substantially eliminate the role of friction or adhesion with the solid surface in the impact dynamics, we work in inverse Leidenfrost conditions \cite{Antonini2013}. This is achieved by impacting a drop or a bead  at ambient temperature on a polished silicon wafer (Si-Mat silicon materials) covered with a thin layer of liquid nitrogen ($N_2$) at $ T=-196 ^\circ $C (see Fig. 1a). Expanded polystyrene is used to build a container of dimension $35$ cm $\times 35$ cm that is filled with liquid nitrogen (the depth of the liquid is typically $10$ cm). Plexiglass is used to cover the polystyrene container and forms an enclosed chamber, which is filled with $N_2$ gas so as to minimize humidity and $N_2$ evaporation. The level of liquid $N_2$ in the bath is maintained below the silicon wafer to avoid the boiling droplets of liquid $N_2$ to hover on the wafer. Two holes are drilled in the polystyrene container to make inlets for compressed $N_2$ gas and liquid $N_2$. Before each impact, the silicon wafer is first cleaned by blowing $N_2$ gas and then a thin layer (typical thickness $50$ nm as measured by ellipsometry) of liquid $N_2$ is deposited on the wafer. Liquid and viscoelastic drops are injected from a syringe pump through a needle. The size of the falling drop is dictated by the inner diameter of the nozzle and the equilibrium surface tension of the sample. In order to maintain a constant drop size, needles with different diameters are used to account for the various sample surface tensions. In the case of elastic beads, a needle attached to a syringe via a flexible tube pins the bead by gently sucking air. On ceasing the suction, the bead is released.
Because the drop or bead is  much warmer than liquid $N_2$, upon impact a vapor cushion forms at the liquid interface due to the evaporation of $N_2$, providing a unique scenario of non wetting and slip conditions that eliminate viscous dissipation~\cite{Chen:2016fo,Antonini2013}. 

We perform impact experiments using three classes of materials: ultrasoft beads of crosslinked gels, liquid drops, and drops of viscoelastic fluids, all beads and drops sharing a fixed diameter $d_0=3.7$ mm. Water (surface tension, $\gamma=72$ mN/m) and mixtures of water and ethanol, with ethanol molar fractions  $0.033$ ($\gamma=50$ mN/m) and  $0.17$ ($\gamma=32$ mN/M) are used as Newtonian liquids.
Polyacrylamide gels are prepared by copolymerization of acrylamide as monomer and
 methylenebisacrylamide as comonomer in the presence of tetramethylenediamine ($0.6$ g/L)
 and sodium persulfate ($0.93$ g/L) as initiators in water~\cite{Menter2000}. Solutions of monomer and
comonomer are mixed in a beaker prior to the addition of the initiators.
The solution is quickly swirled and $26.5$ $\mu$L (corresponding to a drop diameter of
$3.7$ mm) of the solution is transferred immediately to an Eppendorf tube,
filled with poly(methylhydrosiloxane) oil. This oil has a density nearly equal to that of
water ($1.006$ g/ml at $25^{\circ}$C) allowing the drop to float drop, while slowly polymerizing. After the completion of polymerization (typically after $80$ min), the gel bead is taken out of the oil using a pipette and wiped to remove oil from their surfaces.
Sample elasticity is tuned by varying the concentrations of monomer and comonomer.
The shear modulus of the gel, $G_0$, is measured using an indentation technique on bulk pieces of gel. In brief, a rigid sphere is indented in the gel fully covered with pure water and the force along with the indentation depth
is measured. $G_0$ varies between $11$ and $740$ Pa.
We use as viscoelastic samples surfactant-stabilized oil droplets (microemulsions) of diameter $12$ nm, suspended in water and reversibly linked by telechelic polymers. The average number of telechelic stickers per oil droplet is $4$, the mass fraction of oil
droplets $\phi$ varies in the range $(1-3)$ \%. A detailed description is  provided in Ref.~\cite{Arora:2016bu}. The samples are viscoelastic Maxwell fluids, characterized by an elastic plateau $G_0$ and a unique relaxation time $\tau$,  determined by shear rheology.  $\tau$ ranges between $0.1$ and $4$ s, and $G_0$ ranges between $2$ and $21$ Pa.  Because the relaxation time of the viscoelastic fluids is much larger the typical duration of an impact ($\sim10$ ms), for impact experiments, the viscoelastic drops can be considered as elastic beads (large Deborah number). The deformations of these viscoelastic drops are therefore expected to follow those of elastic beads having the same diameter, shear modulus and surface tension. We assume that the surface tension of the polyacrylamide bead is equal to that of pure water, and that of the viscoelastic fluid is equal to that of a bare liquid microemulsion ($\gamma=28$ mN/m)~\cite{Tabuteau:2009ib}.


 \begin{figure}
\includegraphics[width=8.5cm]{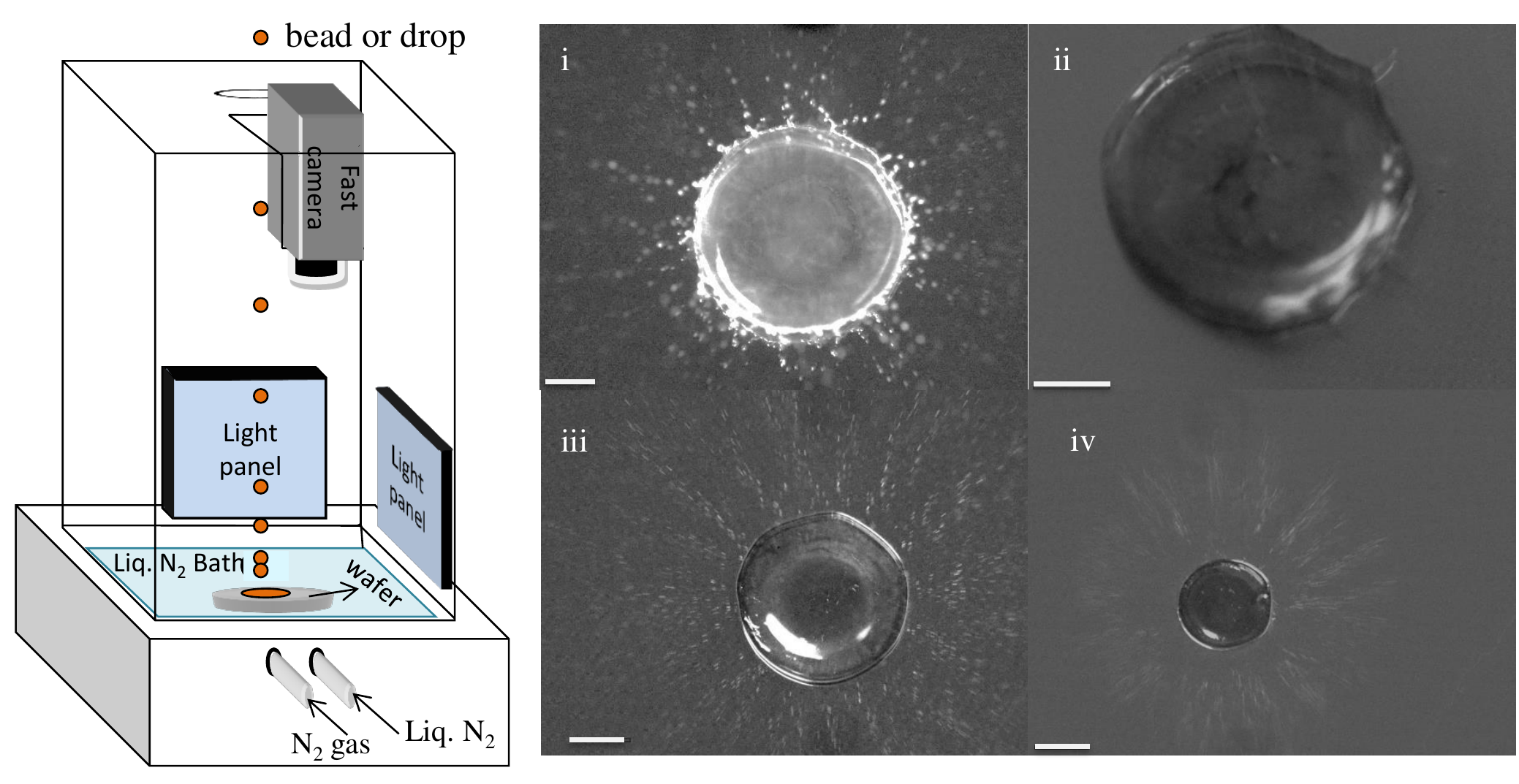}
\caption{left) Experimental set-up; right) Snapshots of liquid, viscoelastic and solid samples at maximum expansion after impact. (i)  Ethanol/water mixture with surface tension  $\gamma=50$ mN/m and impact velocity $v_0=4.35$ m/s; (ii) viscoelastic fluid with shear modulus $G_0=10$ Pa, $\gamma=50$ mN/m, $v_0=3.8$ m/s; (iii, respectively iv) elastic beads with $G_0=35$ Pa (respectively $334$ Pa) and $v_0=4.35$ m/s. Scale bars: $6$ mm.}
\label{figure:1}
\end{figure}

Time series images are recorded after impact using a high-speed camera Phantom V7.3 operated at a rate of $6700$ frames/s. The impact velocity $v_0$  is varied in the range $(1-5)$ m/s by changing the height at which the drop or bead is released. In all cases, the drop or bead expands radially up to a maximal diameter $d_{max}$ and then recedes and rebounds. Figure 1 shows typical snapshots of liquid, solid and viscoelastic sheets taken at their maximal expansion.

Let one consider first an elastic bead.  During its spread, the bead undergoes a biaxial deformation that is quantified at each time by a characteristic  stretching ratio,  $\lambda=d/d_0$, with $d$ the diameter of the sheet, yielding a stored bulk elastic energy $E_{bulk}$
\begin{equation}
\label{eqn:E_bulk}
E_{bulk}\sim \dfrac{1}{2}\dfrac{\pi d_0^3}{6}G_0 (2 \lambda^2 + \dfrac{1}{\lambda^4 }-3)
\end{equation}
For a large maximal spreading factor ($\lambda_{max}\gg1$), the bulk elastic energy at maximal expansion simplifies to  $\text{E}^{max}_{bulk} \sim \dfrac{\pi d_0^3}{6}G_0 \lambda_{max}^2 $. Balancing this energy with the kinetic energy at impact $E_{k}=\dfrac{1}{2}\rho\dfrac{\pi}{6} d_0^3 v_0^2$ leads to the simple scaling  $\lambda_{max} \propto \dfrac{v_0}{U_s}=M$. Figure 2a shows the variation of $\lambda_{max} $ with $M$ for elastic beads with shear moduli varying over almost two orders of magnitude. Although experimental data are in very good agreement with the simple theoretical expectation ($\lambda_{max}\propto M$) for rather stiff samples, they clearly deviate for soft beads ($G_0$ typically smaller than $60$ Pa). Interestingly, deviations occur when the elastocapillary length  $l_{ec}$  (with $\gamma=72$ mN/m, the  surface tension of the gel constituted mainly of water) is larger than the diameter of the beads  $d_0$~\cite{Mora:2013he}, indicating  that the surface elasticity  $\textit{E}^{max}_{surf}\sim \dfrac{1}{2} \pi \gamma \lambda_{max}^2 d_0^2$ dominates over the  bulk elastic energy $\textit{E}^{max}_{bulk}$. Thus, adding the surface energy at maximum expansion in the energy balance ($E_{k}\approx \textit{E}^{max}_{bulk}+\textit{E}^{max}_{surf}$) leads to:
 \begin{equation}
\label{eqn:lambda}
\lambda_{max}\approx\dfrac{1}{\sqrt{2}}\dfrac{v_0}{\sqrt{U_L^2+U_S^2}}=\dfrac{1}{\sqrt{2}}\dfrac{v_0}{U^{\star}}
\end{equation}
Here $U_L= \sqrt{\dfrac{3 \gamma}{\rho d_0}}$ is the typical velocity of free oscillations of a drop \cite{Richard2002}. We thus define a new characteristic  velocity of the material for generalized elastic deformations as $U^{\star} \equiv \sqrt{U_L^2+U_S^2}$.

 Equation~\ref{eqn:lambda} can be alternatively expressed in terms of the elastocapillarity length $l_{ec} $: $\lambda_{max}\approx \dfrac{1}{\sqrt{v_0}}\dfrac{v_0}{U_L}[1+(d_0/l_{ec})]^{-1/2}$. If $U_S \gg U_L$ ($ d_0\ll l_{ec})$, surface tension effects can be neglected and the expansion of an elastic bead is uniquely dominated by its elastic modulus, as observed by Tanaka $et$ $al.$~\cite{Tanaka2003}. On the other hand, when $U_S \ll U_L$ ($d_0\gg l_{ec}$), the bulk elasticity is negligible in comparison to surface elasticity, thus recovering the predictions of Richard $et$ $al.$~\cite{Richard2002} for a simple liquid.

\begin{figure}

\includegraphics[width=8.5cm]{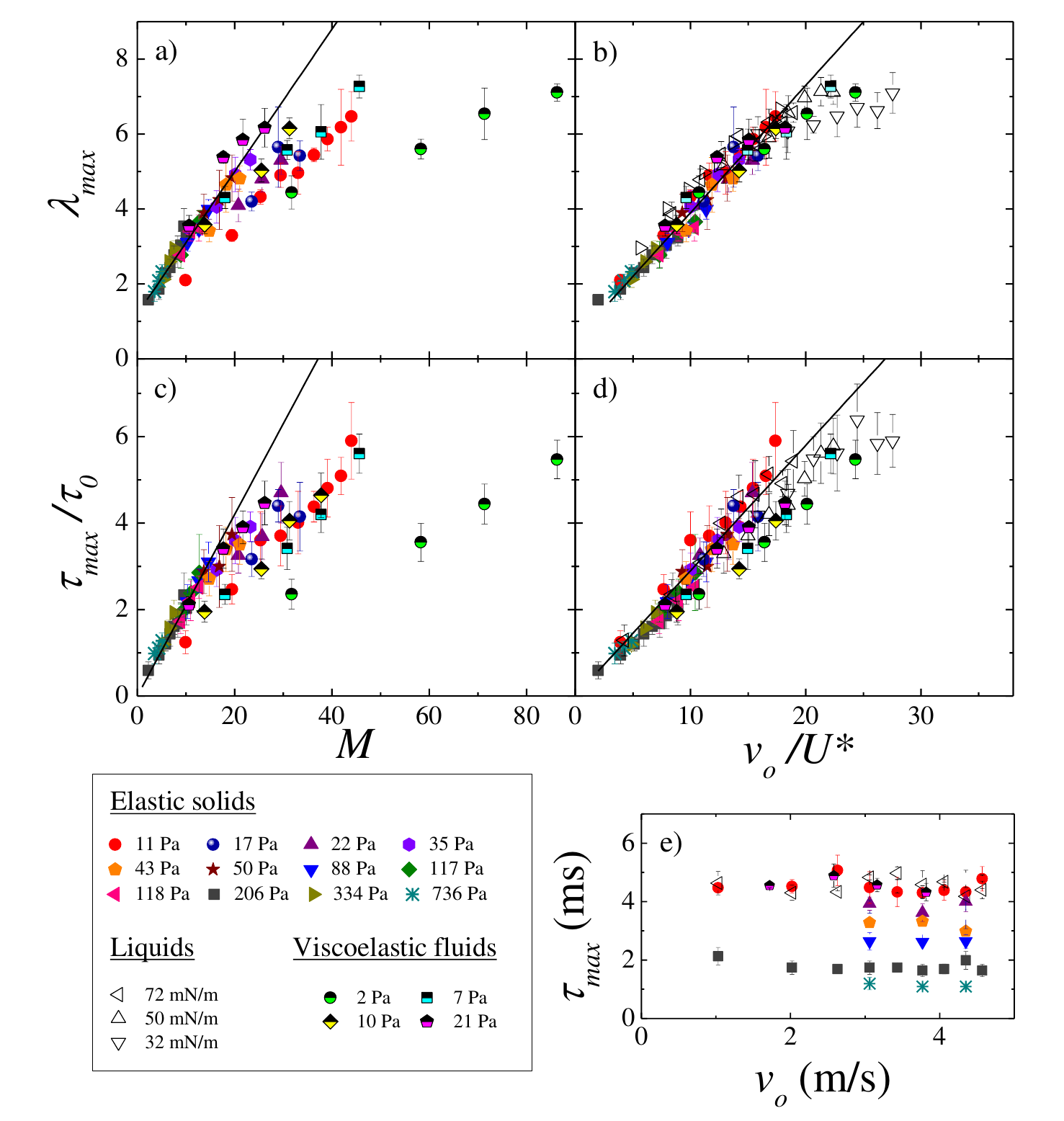}

\caption{Maximal spread parameter, a)  for elastic  beads as a function of the Mach Number, b) for elastic beads, simple liquids and viscoelastic fluids as a function of the impact velocity $v_0$ rescaled by the velocity of generalized elastic deformations $U^{\star}$.  Time at maximum deformation $\tau_{max}$ divided by the impact time $\tau_0$, c) for elastic beads as a function of the Mach Number, d) for elastic beads, simple liquids and viscoelastic fluids as a function of $v_0/U^{\star}$.  Each experiment has been repeated five times. Error bars correspond to $\pm$ standard deviation. e) $\tau_{max}$ as a function of $v_0$ for elastic beads and liquid drops. The symbols are the same for all plots.}
\label{figure:2}
\end{figure}

To confront the theoretical scaling of Eq.~\ref{eqn:lambda} with experiments, $\lambda_{max}$ is plotted against $v_0/U^{\star}$ in Fig.~2b. We find that all data acquired for elastic beads (plain symbols) collapse onto a master curve exhibiting a perfect linear variation, whatever the value of the elastic modulus is. Although Eq.~\ref{eqn:lambda} should in principle be valid only in the case of very large deformations ($\lambda_{max}\gg1$)
 we find that the asymptotic linear relation describes the experimental results very well even for moderate deformations ($\lambda_{max} \approx 2$).  Notably, we find that data acquired using viscoelastic drops (half-plain symbols in Fig. 2b), and Newtonian liquid drops for which $U^\star= U_L$ ($U_S=0$) (empty symbols in Fig. 2b), collapse on the same master curve. In spite of a neat universal scaling, data tend however to deviate from the linear prediction at high impact velocities for simple liquids and viscoelastic fluids due to splashing, leading to a loss of volume~\cite{Mundo1995, Rioboo:2002gk}, or inertial dissipations in the rim~\cite{Clanet2004}. Overall, our results establish a universal scaling of the maximum deformation of elastic beads, viscoelastic and liquid drops, under the conditions of negligible viscous dissipation provided the bulk and surface elasticity are correctly taken into account.

We measure the time evolution of the sheet diameter $d$ for the three classes of samples (elastic, viscoelastic and liquid). For the sake of clarity we just show data at a fixed impact velocity ($v_0=4.35$ m/s) for elastic beads with varying stiffness (Fig. 3a). All curves show similar features of expansion and retraction, the maximum diameter of the sheet being reached earlier in times with increasing elasticity. We model the spreading dynamics as a one-dimensional  ($1$D) harmonic oscillator, as previously done independently for Newtonian drops~\cite{Biance2006,  Andrew:2017do} and elastic beads~\cite{Tanaka:2006gp}. Conservation of total (elastic and kinetic) energy reads in the limit of large deformation ($\lambda\gg1$):
\begin{equation}
\label{eqn:dynamics1}
\dfrac{1}{2}m\dot{d}^2+\dfrac{1}{2}kd^2=\dfrac{1}{2}kd_{max}^2
\end{equation}
where $d$ is the diameter of the sheet at time $t$, $m=\rho \pi d_0^3/6$ is the mass of the sheet equal to that of the impacting object, $k=\pi \gamma+\pi\dfrac {d_0}{3} G_0 $ is the spring constant that combines  the bulk and surface elastic contributions. The time elapsed since impact to reach maximum expansion, $\tau_{max}$, is then the quarter of the period of oscillation:
$\tau_{max}\approx\dfrac{\pi}{2\sqrt{2}} \dfrac{d_0}{U^\star}$. Note that $\tau_{max}$ is half the rebound time $\tau_R$, but is much easier to measure. Interestingly, we measure (Fig. 2e) that $\tau_{max}$ is independent of the impact velocity in accordance with the $1D$ harmonic oscillator prediction. Once $\tau_{max}$ is rescaled with the collision time $\tau_0=d_0/v_0 $ a similar dependance of $\tau_{max}/\tau_ 0$ with the reduced impact velocity  $v_0/ U^\star$ as  for the maximum spread parameter $\lambda_{max}$ (Eq.~\ref{eqn:lambda}) is  recovered:
\begin{equation}
\label{eqn:taumax}
\dfrac{\tau_{max}}{\tau_0}\approx\ \dfrac{\pi}{2\sqrt{2}} \dfrac{v_0}{U^\star}
\end{equation}

\begin{figure}
\includegraphics[width=8.5cm] {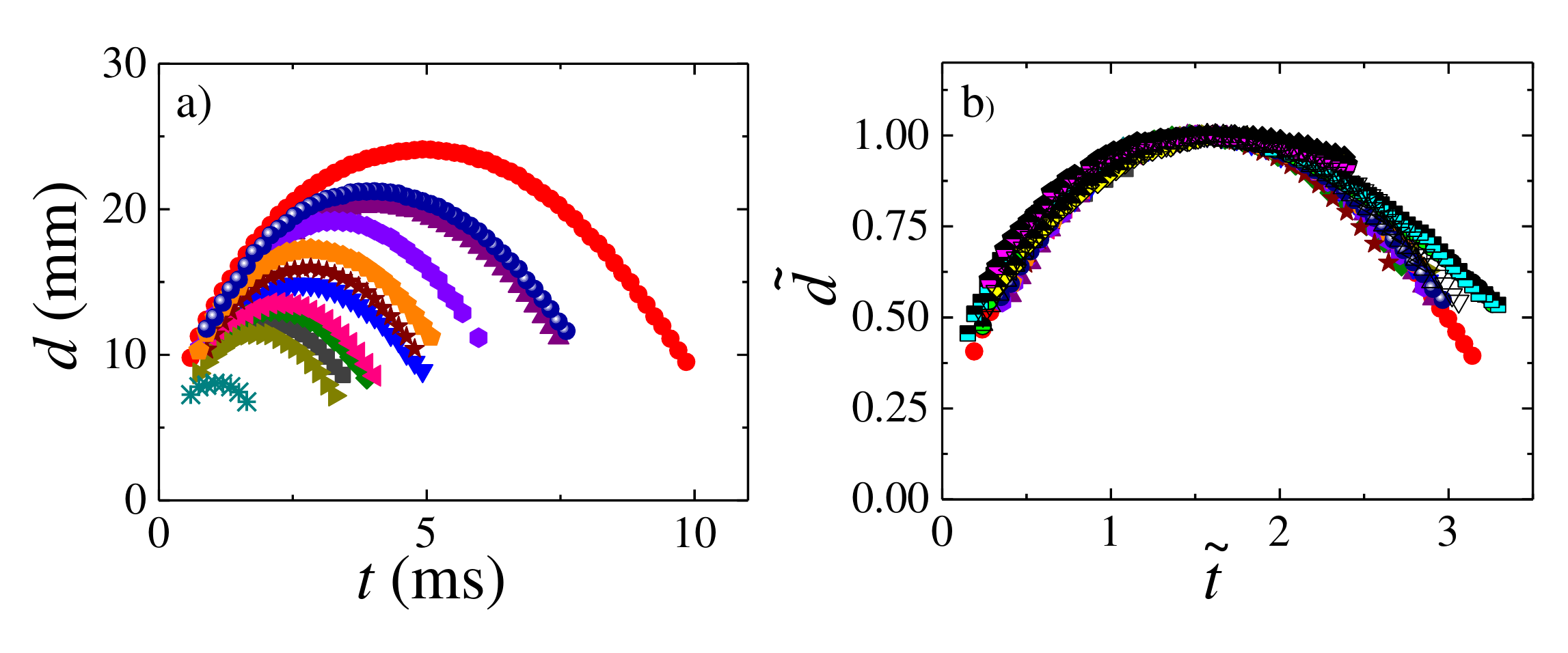}
\caption{ a) Time evolution of the sheet diameter for elastic beads with different stiffness. The impact velocity is $v_0=4.35 \mathrm{m/s}$. The symbols are the same as in Fig. 2b) Same data as in a) plotted in rescaled units (see text). Data for viscoelastic and liquid drops are also shown. The symbols are the same as in Fig. 2.}
\label{figure:3}
\end{figure}

While for the softest beads, experimental data $\tau_{max}$ departs from a linear dependence with $M$ (see Fig. 2c), as expected if surface effects are negligible~\cite{Tanaka:2006gp}, they nicely follow the theoretical predictions of Eq.~\ref{eqn:taumax} for all explored elastic moduli and impact velocities (see Fig. 2d), confirming the crucial importance of elastocapillary effects. Experimental results for simple liquids and viscoelastic fluids also merge on the same master curve, with a deviation from the theoretical linear variation at high impact velocity for the liquid samples due to the loss of mass induced by splashing. 
The unified universal behavior of the impact of elastic beads, viscoelastic and  Newtonian drops is also confirmed for the dynamics of the sheet, that is predominantly a simple harmonic motion driven by surface tension and bulk elastic energy. Using adimensional units $\tilde{d}= d/d_{max}$ and $\tilde{t}=\omega t$ with  $\omega=\sqrt{\frac{k}{m}}$, Eq.(\ref{eqn:dynamics1}) reduces to the adimensional equation  $\dot{\tilde{d}}^2+\tilde{d}^2=1$. Using the same rescaling for the experimental data, a nice collapse of all data sets (Fig. 3b), corresponding to different bulk and/or surface  elastic  properties of the impacting objects  is observed  at least  for the expansion regime ($t\leqslant( \tau$).  Weak deviations from this simple general behavior occurs in the retraction regime as already observed for elastic beads \cite{Tanaka:2006gp} or Newtonian drops \cite{Biance2006}, which originates  in terms of the existence of a rim, or drop break-up or departure from a cylindrical symmetry.\cite{Andrew:2017do}.  

In conclusion, we have highlighted  the importance of elastocapillarity to properly describe the physics of impact. When viscous and solid  friction dissipations are negligible, the spreading dynamics caused by an impact can be described with a unique scaling law with a  characteristic spreading velocity that includes both surface and bulk elasticity. We have experimentally demonstrated the validity of the scaling, whatever the nature of the impacting object, soft elastic beads, viscoelastic drops, liquid drops.

\begin{acknowledgments}
This work was supported by the EU (Marie Sklodowska-Curie ITN Supolen, Grant No.
607937).
We thank Antonio Stocco for his help in ellipsometry measurements.
\end{acknowledgments}


\bibliographystyle{aipnum4-1}
\bibliography{Srishti}


\end{document}